\documentclass[aps,prl,twocolumn,groupedaddress,amsmath,amssymb,showpacs]{revtex4}

\usepackage{graphicx}
\usepackage{dcolumn}
\usepackage{bm}

\begin{document}

\title{Large linear magnetoresistance in the Dirac semimetal TlBiSSe}
\author{Mario Novak}
\email{mnovak@sanken.osaka-u.ac.jp}
\author{Satoshi Sasaki}
\author{Kouji Segawa}
\author{Yoichi Ando}
\email{y_ando@sanken.osaka-u.ac.jp}
\affiliation{Institute of Scientific and Industrial Research, Osaka University, 
Ibaraki, Osaka 567-0047, Japan}

\date{\today}

\begin{abstract}

The mixed-chalcogenide compound TlBiSSe realizes a three-dimensional
(3D) Dirac semimetal state. In clean, low-carrier-density single
crystals of this material, we found Shubnikov-de Haas oscillations to
signify its 3D Dirac nature. Moreover, we observed very large linear
magnetoresistance (MR) approaching 10,000\% in 14 T at 1.8 K, which
diminishes rapidly above 30 K. Our analysis of the magnetotransport data
points to the possibility that the linear MR is fundamentally governed
by the Hall field; although such a situation has been predicted for
highly-inhomogeneous systems, inhomogeneity does not seem to play an
important role in TlBiSSe. Hence, the mechanism of large linear MR is an
intriguing open question in a clean 3D Dirac system.

\end{abstract}






\maketitle

The discoveries of graphene \cite{Geim} and three-dimensional (3D)
topological insulators \cite{Hasan-Kane, Qi-Zhang, Ando} greatly
advanced the physics of two-dimensional (2D) massless Dirac fermions. In
comparison, 3D massless Dirac fermions, whose Hamiltonian involves all
three Pauli matrices, have attracted much less attention. This is due
partly to the shortage of concrete materials to give access to the
massless Dirac physics in 3D, although {\it massive} 3D Dirac fermions
in Bi are long known to present interesting physics
\cite{Fukuyama, OngBi, Behnia}. However, this
situation has changed recently, and materials to realize 3D massless
Dirac fermions are currently attracting significant attention because of
the interest in a new type of topological materials called Weyl
semimetals \cite{XWan11, Vishwanath}. In recent literature, materials
realizing spin-degenerate 3D massless Dirac fermions are called ``3D
Dirac semimetal", while those realizing a pair of spin-nondegenerate 3D
massless fermions are called ``Weyl semimetal"; the latter is derived
from the former by breaking time-reversal symmetry or space-inversion
symmetry (or both) to split the spin-degenerate Dirac cone into two
spin-nondegenerate ones \cite{Vishwanath}.

Recently, the 3D Dirac semimetal phase has been shown to exist in
Na$_3$Bi \cite{Wang12-2, Liu14, Xu14} and Cd$_3$As$_2$ \cite{Wang13-2,
Neupane14, Borisenko14}. Also, such a phase is known to exist at the
topological phase transition point of TlBi(S$_{1-x}$Se$_x$)$_2$
\cite{Sato11, Xu11-2, Burkov11, Singh12}, Pb$_{1-x}$Sn$_x$Se
\cite{Zeljkovic14}, Bi$_{1-x}$Sb$_x$ \cite{Teo08} etc., where the bulk
band gap necessarily closes. In those materials, the Weyl semimetal
phase would be realized by magnetic doping, breaking the crystal
inversion symmetry, or applying external magnetic field \cite{Burkov11,
Singh12, Kim13, Bulmash14}. Besides being potential parent materials of
Weyl semimetals, the Dirac semimetals offer a new playground to explore
the physics of massless Dirac fermions in larger spatial degrees of
freedom than the 2D case, which may change the characteristic transport
properties in a nontrivial way. 

In this Letter, we report our magnetotransport studies of TlBiSSe, where
the 3D Dirac semimetal phase is realized as a result of the topological
phase transition between the topological insulator (TI) TlBiSe$_2$ and an
ordinary insulator TlBiS$_2$ \cite{Sato11, Xu11-2}. In TlBiSSe, as the
Fermi level is tuned close to the Dirac point, the magnetoresistance
(MR) grows very rapidly, and its magnetic-field dependence is found to
become linear in high magnetic fields. Surprisingly, in samples with the
Fermi energy $E_F$ of about 20 meV, we observed very large linear MR
approaching 10,000\% at 14 T. Our analysis of the magnetotransport data
strongly suggests that the linear MR is somehow governed by the Hall
field, but its origin is not explicable with existing theories for
linear MR, pointing to new physics peculiar to 3D Dirac fermions.

The TI nature of TlBiSe$_2$ was found in 2010 \cite{Sato10, Kuroda10,
Chen10}, and the existence of the topological phase transition in
TlBi(S$_{1-x}$Se$_x$)$_2$ was discovered in 2011 \cite{Sato11, Xu11-2}:
according to the angle-resolved photoemission spectroscopy (ARPES) data,
the bulk band gap in TlBi(S$_{1-x}$Se$_x$)$_2$ closes at $x$ = 0.5,
across which the band inversion takes place and the $x > 0.5$ side
obtains nontrivial $Z_2$ topology signified by the appearance of
topological surface states. This means that the zero-gap semimetallic
state is realized in TlBiSSe, in which S and Se occupy the chalcogen
site in a mixed way. Single crystals of TlBi(S$_{1-x}$Se$_x$)$_2$ grown
from stoichiometric melts are always $n$-type with the typical carrier
density of 10$^{20}$ cm$^{-3}$ \cite{Sato10}. Motivated by a recent
report \cite{Kuroda13}, we have grown crystals of TlBiSSe with a Tl-rich
starting composition \cite{SM}, and succeeded in reducing the bulk
carrier density down to 10$^{17}$ cm$^{-3}$ level. High crystallinity of
our single crystals is confirmed by x-ray diffraction (XRD) analysis
[Fig. 1(a)] and Laue analysis. Although the crystals are grown from
off-stoichiometric melts, inductively-coupled plasma atomic-emission
spectroscopy (ICP-AES) analysis confirmed that the compositions of the
grown crystals are close to stoichiometry, and electron-probe
microanalysis (EPMA) data assured that there is no segregation of
constituent elements, as discussed in the Supplemental Material
\cite{SM}. Experimental details of our transport measurements are also
described in \cite{SM}.

The temperature dependencies of the in-plane resistivity,
$\rho_{xx}(T)$, of three representative TlBiSSe samples with
significantly different carrier densities are shown in Fig. 1(b); note
that the vertical axis is in logarithmic scale, and the residual
resistivity ratio of the lowest-carrier-density sample (S3) is as large
as 73. We have actually measured many more samples than are shown in
Fig. 1(b), and Fig. 1(c) shows that the low-temperature transport
mobility $\mu_{t}$ (assessed from $\rho_{xx}$ at 1.8 K and the carrier
density $n$) increases systematically with decreasing $n$; for example,
$\mu_{t}$ increases by 110 times between samples S1 ($n$ = 8.8 $\times$
10$^{19}$ cm$^{-3}$) and S3 ($n$ = 3.8 $\times$ 10$^{17}$ cm$^{-3}$).
This is in contrast to the case of 2D Dirac systems like graphene
\cite{ZhangGraphene} and 3D TIs \cite{Yang14}, where $\mu_{t}$ shows an
enhancement only when the Fermi level is tuned very close to the Dirac
point. 

\begin{figure}
\includegraphics[width=8.5cm]{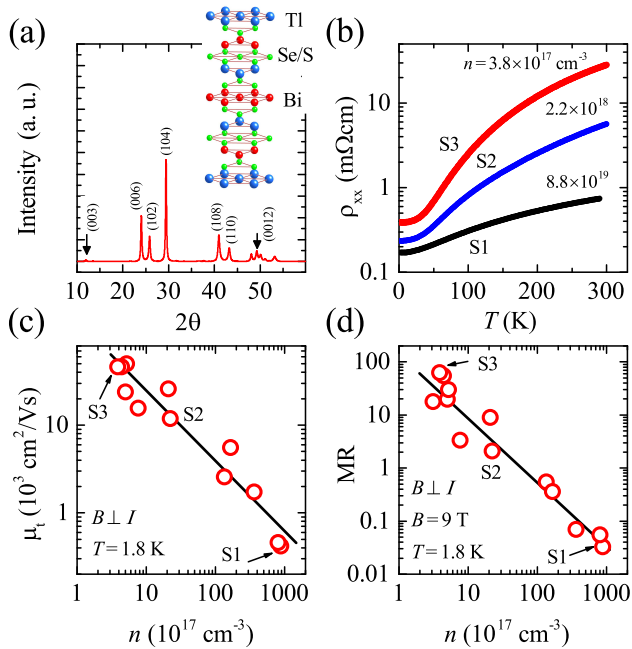}
\caption{
(a) Powder XRD pattern of a typical TlBiSSe crystal; inset shows its 
crystal structure. 
(b) $\rho_{xx}(T)$ behavior of three samples with different carrier
densities. 
(c, d) Plots of low-temperature mobility and the magnitude of MR
vs $n$. The MR values shown here are for 9 T at 1.8 K. 
 }
\label{Fig1}
\end{figure}

Perhaps more surprising is the very rapid increase in MR with decreasing
$n$; for example, the MR at 9 T [Fig. 1(d)] changes by almost 2,000
times between S1 and S3. Here, MR is defined by
$[\rho_{xx}(B)-\rho_{xx}(0 {\rm \,T})]/\rho_{xx}(0 {\rm \,T})$. To gain
insights into the large MR, Fig. 2(a) shows how the MR behavior in
sample S3 changes when the magnetic field is tilted from perpendicular
to parallel directions. The angular dependence is more directly shown in
Fig. 2(b), where the magnitude of the MR in 14 T is plotted as a
function of the angle $\theta$, which is defined in the inset of Fig.
2(a). The dipole-like pattern seen in Fig. 2(b) is well described by the
$\cos \theta$ function (red solid line), meaning that the MR is almost
entirely governed by the perpendicular component of the magnetic field,
even though the present system is 3D. The magnetic-field dependence of
$\rho_{xx}$ at low field is plotted in Fig. 2(c) for $\theta$ =
0$^{\circ}$ and $92^{\circ}$, both of which present the ordinary $B^2$
behavior below $\sim$0.1 T; this suggests that the origin of the linear
MR is different from the famous linear MR in Ag$_{2+\delta}$Se and
Ag$_{2+\delta}$Te \cite{Xu97}, where the linearity is observed from as
low as 1 mT. Note that, due to the high mobility of the sample S3, the
condition $\omega_c \tau_t$ = $\mu_{t} B$ = 1 ($\tau_t$ is the transport
scattering time and $\omega_c = e B/m_c$ is the cyclotron frequency with
$m_c$ the cyclotron mass) is achieved in only 0.2 T, and hence the
standard theory for MR for a closed Fermi surface \cite{Abrikosov88}
would predict a saturation at $B \gg$ 0.2 T; nevertheless, as one can
see in Fig. 2(a), this sample presents non-saturating linear MR above
$\sim$6 T.

\begin{figure}
\includegraphics[width=8.5cm]{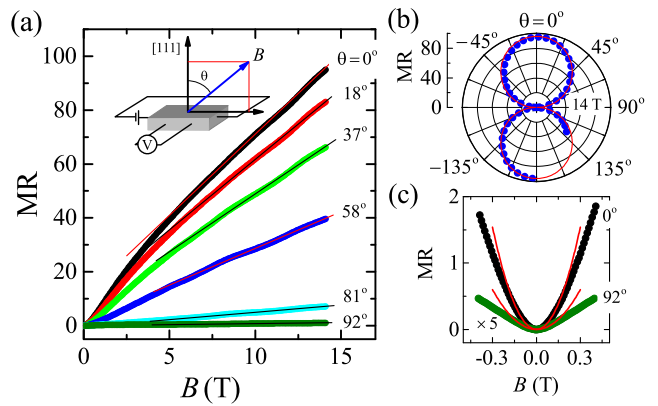}
\caption{MR in sample S3 at 1.8 K.
(a) $\rho_{xx}(B)$ behavior for various magnetic-field
angles form transverse ($\theta =0^{\circ}$, $B$ field in the
[111] direction) to the near-longitudinal ($\theta = 92^{\circ}$)
configurations; inset depicts the definition of $\theta$.
(b) Dipole-like $\theta$ dependence of the magnitude of MR at 14 T,
which follows the $\cos \theta$ dependence (red solid line). 
Due to the restriction of the rotation stage, the range of 
$\theta$ does not span the whole 360$^{\circ}$. 
(c) Low-field MR showing ordinary $B^2$ behavior; the red solid lines
show the fits to the $B^2$ function.}
\label{Fig2}
\end{figure}

The temperature dependence of this linear MR signifies its unique
nature, not reported before for other systems showing large linear MR
\cite{Xu97, Hu08, Wang12, Friedman10, Porter12, Kozlova}.
Figures 3(a)-3(c) show $\rho_{xx}$ vs $B$ at various temperatures, where
one can see that the characteristic field above which the linear MR is
observed remains around 6 T up to 150 K, but at higher temperature the
linear MR disappears. More importantly, the size of MR changes little
between 1.8 and 30 K, but at higher temperature it diminishes rapidly.
This temperature dependence is summarized in Fig. 3(d), where the
dependence of $n$ on temperature is plotted together; one can see that
$n$ changes only by a small amount, and hence the rapid decline in MR
has little to do with the thermal activation of carriers. On the other
hand, as shown in Fig. 3(e), the size of MR depends linearly on $\mu_t$,
implying that the reason for the rapid decline in MR is the phonon
scattering which restricts $\mu_t$ at high temperature. 

It is prudent to mention that, even though the MR is unusual in many
respect, it obeys the Kohler's rule \cite{Abrikosov88} (see \cite{SM}
for details), meaning that $\rho_{xx}$ depends on the magnetic field
only through the form $B \tau_t$ (which is the case in the semiclassical
relaxation-time approximation). In passing, the MR data at 200 and 300 K
can be described by the conventional form $aB^2/(1+bB^2)$ \cite{Pippard89}.

\begin{figure}
\includegraphics[width=8.5cm]{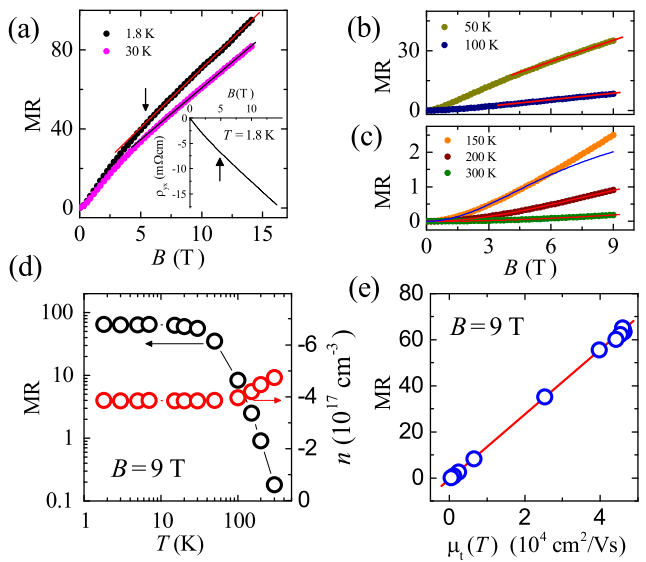}
\caption{MR in sample S3 for $\theta$ = 0$^{\circ}$. 
(a, b, c) $\rho_{xx}(B)$ behavior at various temperatures; note the 
different vertical scales between panels.
The inset in (a) shows the $\rho_{yx}(B)$ behavior at 1.8 K; the arrows
mark the change in slope. The straight lines in (a) and (b) are fits to
the linear part, while the solid lines in (c) are fits of the low-field
part to the classical $aB^2/(1+bB^2)$ law. 
(d) Temperature dependences of the magnitude of MR at 9 T (left axis)
and the carrier density calculated from $R_H$ at each $T$ (right axis). 
(e) Plot of the magnitude of MR at 9 T vs the transport mobility
$\mu_t$, which changes with $T$.
}
\label{Fig3}
\end{figure}

\begin{figure}
\includegraphics[width=8.5cm]{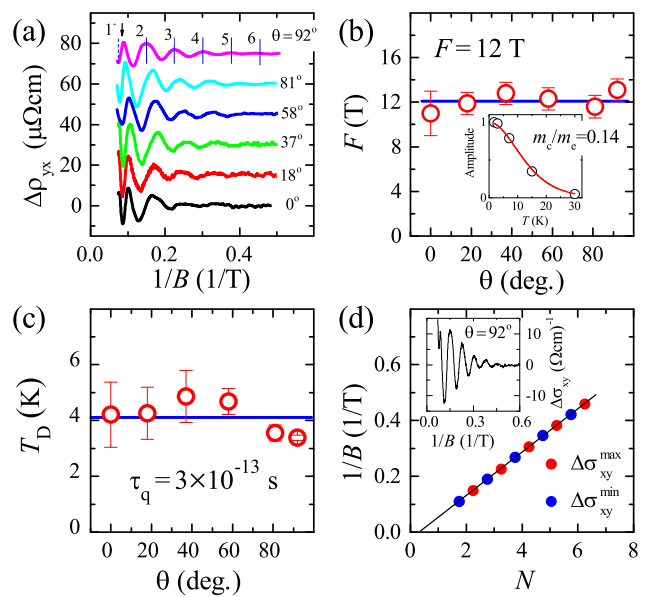}
\caption{SdH oscillations in sample  S3. 
(a) SdH oscillations in $\rho_{yx}$ vs $1/B$ at 1.8 K for various
magnetic-field directions. The equidistant maxima are indicated by
vertical lines, with the exception of the 1st Landau level which shows
spin splitting.
(b) $\theta$ dependence of the oscillation frequency; inset shows the 
temperature dependence of the oscillation amplitude for $\theta$ = 0$^{\circ}$
together with the fitting to the Lifshitz-Kosevich theory which gives
$m_c/m_e$ = 0.14. 
(c) $\theta$ dependence of the Dingle temperature.
(d) Landau-level index plot for oscillations in $\sigma_{xy}$ measured
at 1.8 K and $\theta$ = 0$^{\circ}$; inset shows the oscillations in
$\Delta \sigma_{xy}$ which is obtained by subtracting a smooth
background from $\sigma_{xy}$. Following the principle in Refs.
\cite{Ando, TaskinSnTe} and assuming electron carriers, we assign the
index $N+\frac{1}{4}$ and $N+\frac{3}{4}$ to the maxima and minima in
$\Delta \sigma_{xy}$, respectively. Solid line is a linear fitting to
the data, giving the intercept on the $N$ axis of 0.34. 
}
\label{Fig4}
\end{figure}

The low-carrier-density samples are clean enough to present Shubnikov-de
Haas (SdH) oscillations, which are the source of the wiggles in the MR
data at high $B$. Clear observation of SdH oscillations signifies not
only a high mobility but also a high homogeneity of local carrier
density, since a variation of local carrier density would result in a
spread of SdH frequencies to smear the oscillations. In the case of
TlBiSSe, a larger number of oscillation cycles are discernible in
$\rho_{yx}(B)$ than in $\rho_{xx}(B)$, so we mainly used the former for
the following analysis. Figure 4(a) shows SdH oscillations in
$\rho_{yx}$ for varying magnetic-field angle $\theta$ after removing the
linear background. The Fourier transform gives only one frequency, whose
dependence on $\theta$ is shown in Fig. 4(b); these data reveal a very
small spherical (isotropic) Fermi surface (FS).

The averaged frequency $F$ = 12 T gives the FS radius $k^{\rm 3D}_F =
1.9 \times 10^{6}$ cm$^{-1}$ and the carrier density $n_{\rm SdH} = 2.4
\times 10^{17}$ cm$^{-3}$ \cite{note3}. From the temperature dependence
of the oscillation amplitude at $\theta = 0^{\circ}$ [Fig. 4(b) inset],
we obtain $m_c = 0.14 m_e$ ($m_e$ is the free electron mass) by using
the Lifshitz-Kosevich (LK) theory \cite{Shoenberg}. This allows us to
determine the Dingle temperature $T_{\rm D}$, which is plotted in Fig.
4(c) as a function of $\theta$. Its average value, $T_{\rm D}$ = 4.1 K,
gives the quantum scattering time $\tau_q = \hbar/(2\pi k_B T_{\rm D}) =
3 \times 10^{-13}$ s. This is to be compared with the transport
scattering time $\tau_t = 3.7 \times 10^{-12}$ s assessed from $\mu_t$;
the difference, which in this case is about 10 times, is usually
associated with the difference in the rates between forward and backward
scatterings \cite{DasSarma}; obviously, small-angle (forward)
scatterings are relatively strong in TlBiSSe, which happens when 
weak scattering potentials predominate. Other parameters of interest are
obtained as follows: the quantum mobility $\mu_{q} \equiv e \tau_q / m_c
\approx$ 3500 cm$^2$/Vs, Fermi velocity $v_F = \hbar k_F / m_c = 1.6
\times 10^5$ m/s, and the Fermi energy (measured from the Dirac point)
$E_F = \hbar v_F k_F$ = 20 meV. 

An important information derived from SdH oscillations is the Berry
phase \cite{Ando, Kopelevich, Murakawa}. We made the Landau-level (LL)
index plot based on the positions of minima and maxima in $\sigma_{xy}$
\cite{TaskinSnTe} as a function of $1/B$ [Fig. 4(d) inset]. In a system
with 3D FS, the intercept of the index plot on the $N$ axis is expected
to be $0 \pm 1/8$ for Schr\"odinger fermions, while it should be $1/2
\pm 1/8$ for Dirac fermions (the sign before 1/8 should be + for holes
and $-$ for electrons) \cite{Kopelevich, Murakawa}. In our case, the
intercept is 0.34 [Fig. 4(d)], which is close to $1/2 - 1/8$ = 0.375 and
hence is consistent with 3D Dirac electrons.

We now discuss the possible mechanism of the observed large linear MR.
There are several theoretical models which predict linear MR for
low-carrier-density systems. Abrikosov \cite{Abrikosov98, Abrikosov00}
proposed a quantum interpretation of the phenomena by assuming a gapless
linear dispersion and the system to be in the ultra-quantum limit. The
main feature of this model, apart from the linear MR, is the stability
against temperature; this is because the condensation to the lowest LL
is robust until level broadening causes an overlap of adjacent LLs.
TlBiSSe is a gapless 3D Dirac system, and thus the Abrikosov model of
quantum linear MR would be appropriate for describing the observed linear
MR. In this regard, the linear MR in the transverse orientation
($\theta$ = 0$^{\circ}$) sets in at $\sim$6 T, which corresponds to the
situation when the Fermi level is in the 2nd LL; such a situation was
previously argued to be sufficiently close to the ultra-quantum limit to
observe the quantum MR \cite{Hu08, Wang12, Friedman10}. However, the
strong decrease of the MR above $\sim$30 K [Fig. 3(d)] contradicts the
Abrikosov's model.

Thus we turn to other models which can predict linear non-saturating MR
in a system with small 3D FS. A classical one is by Herring
\cite{Herring60}, who developed a perturbation theory for a system with
weak inhomogeneity in the carrier density and showed that the
fluctuations in the Hall field due to the inhomogeneity will lead to
linear MR. Parish and Littlewood (PL) \cite{Parish03, Hu07} proposed a
more comprehensive model which is valid also in the strong inhomogeneity
limit and showed that the inhomogeneity will cause distortions in the
current paths, which in turn causes the Hall field to contribute to the
MR in a symmetric manner with respect to $\pm B$. In this respect, the
$\theta$ dependence of the MR [Fig. 2(b)], which suggests that only the
perpendicular component of the magnetic field is responsible for the MR,
seems to support the scenario that the linear MR originates from the
Hall field. Moreover, the data for in-plane magnetic field rotation
(described in the Supplemental Material \cite{SM}) are also consistent
with this scenario. In addition, it is suggestive that a change in slope
of $\rho_{xx}(B)$ that occurs at around 5 T seems to be correlated with
a similar change in slope of $\rho_{yx}(B)$ at the same field [Fig.
3(a)].

An important clue comes from the Hall angle $\theta_{\rm H}$. According
to the semiclassical theory for a single-band metal, the relation $\tan
\theta_{\rm H} = \rho_{yx}/\rho_{xx} = \sigma_{xy}/\sigma_{xx} =
\omega_c \tau_t$ should hold. However, if we calculate these values for
the sample S3 in 14 T, $\tan \theta_H = \rho_{yx}/\rho_{xx}$ = 0.5,
whereas $\omega_c \tau_t = \mu_t B$ = 65. Therefore, there is a
two-orders-of-magnitude difference between what is purported to be the
same parameter. This is significant, and it strongly supports the
scenario that MR is actually governed by the Hall field rather
than the scattering.

In the PL and Herring's model, the existence of inhomogeneity is
essential. However, in our samples good crystallinity was confirmed by
XRD and Laue analysis, and EPMA data confirmed that there is no
segregation of constituent elements \cite{SM}. Also, the average donor
distance $l_{\rm imp} \simeq n^{-1/3}_{\rm SdH}$ = 15 nm and the Debye
screening length $l_{\rm Debye}$ = 3 nm \cite{note5} are both short;
thus, the low temperature mean free path $\ell = v_F \tau_t$ = 600 nm
does not support the impurities to be the source of strong
inhomogeneity. Whilst the linear relation between MR and $\mu_t$ [Fig.
3(e)] is along the lines with the prediction of PL model, the decline of
the mobility in this case is due to phonon scattering and is not related
to inhomogeneity. Therefore, while the essential spirit of the PL model
is valid and the Hall field appears to be the fundamental source of the
linear MR, the actual mechanism to bring about such a situation is
obviously an open question. 

Finally, we mention that in a recent wok on another 3D Dirac system,
Cd$_3$As$_2$ \cite{Liang14}, a gigantic MR was observed in very clean
samples (with $\mu_t > 10^7$ cm$^2$/Vs) as a result of lifting of the
strong protection from backscattering, which was reflected in the fact
that $\mu_t$ was 10$^4$ times larger than $\mu_q$; this is different
from our situation. In more disordered samples of Cd$_3$As$_2$ with
$\mu_t \simeq 10^4$ cm$^2$/Vs, a large linear MR was observed, but it
starts from very low field and it persists to 300 K, both of which
suggests that it is more in line with the PL model than the TlBiSSe
case. Note that an important difference between Cd$_3$As$_2$ and TlBiSSe
is the number of Dirac cones (two vs one), and the former has a peculiar
anisotropy and an additional valley degrees of freedom.

In summary, we found that in the 3D Dirac semimetal TlBiSSe, a reduction
in carrier density $n$ leads to a rapid increase in the transport
mobility $\mu_t$ and transverse magnetoresistance (MR). In samples with
$n \simeq 10^{17}$ cm$^{-3}$, $\mu_t$ becomes 5 $\times$ 10$^4$
cm$^2$/Vs and linear MR whose magnitude reaches almost 10,000\% in 14 T
was observed at 1.8 K. This linear MR is governed by the perpendicular
component of the magnetic field, and the large discrepancy between $\tan
\theta_H$ and $\omega_c \tau_t$ points to the scenario that the Hall
field is the fundamental source of the linear MR. Nevertheless,
inhomogeneity does not seem to play an important role here, and the
exact mechanism to produce the large liner MR is yet to be determined.

We thank R. Sato for help in crystal growths, and A. A. Taskin for
helpful discussions. This work was supported by JSPS (KAKENHI 
24540320 and 25220708), MEXT (Innovative Area ``Topological Quantum
Phenomena" KAKENHI), AFOSR (AOARD 124038), Inamori Foundation, and
the Murata Science Foundation.


\begin{thebibliography}{50}
\bibitem{note3}
The carrier density determined from the low-field $R_H$ for the sample S3 is 
$n = 3.8 \times 10^{17}$ cm$^{-3}$.
\bibitem{note5}
We calculated the Debye screening length from $l_{\rm Debye} = [\epsilon
k_B T/ (n e^2) ]^{1/2}$ with the dielectric constant $\epsilon \approx$
20 and $T$ = 1.8 K.
\end{thebibliography}

\end{document}